\documentclass[sigconf]{acmart}

\usepackage{CJK}
\AtBeginDocument{%
  }

\setcopyright{acmlicensed}
\copyrightyear{2026}
\acmYear{2026}
\acmDOI{XXXXXXX.XXXXXXX}
\acmConference[DIS '26]{ACM Designing Interactive Systems Conference}{June 13--17,
  2026}{Singapore}
\acmISBN{978-1-4503-XXXX-X/2018/06}

\usepackage{multirow}
\usepackage{tikz}

\usepackage{listings}
\usepackage{xcolor}

\definecolor{varbg}{HTML}{E8EAED} 
\definecolor{promptbg}{HTML}{F8F9FA}
\definecolor{promptframe}{HTML}{D1D5DB}

\usepackage{calc} % 建议加上

\lstdefinestyle{beautiful_listing}{
    basicstyle=\ttfamily\small\linespread{0.92}\selectfont,
    backgroundcolor=\color{promptbg},
    rulecolor=\color{promptframe},
    frame=single,
    framesep=8pt,
    framerule=0.5pt,
    breaklines=true,
    breakatwhitespace=true,
    columns=fullflexible,
    keepspaces=true,
    showstringspaces=false,
    numbers=none,
    escapeinside={(*@}{@*)},
    linewidth=\dimexpr\linewidth-8pt-1pt\relax,
    xleftmargin=0pt,
    xrightmargin=0pt,
    framexleftmargin=0pt,
    framexrightmargin=0pt
}

\newcommand{\revision}[1]{\leavevmode{\textcolor[RGB]{220, 20, 60}{#1}}}
\newcommand{\remark}[1]{\textcolor[RGB]{150, 200, 0}{#1}}

\newcommand{\removed}[1]{\leavevmode{\color{red}{\st{#1}}}}

\def \cleanversion{} % remove this comment to generate a clean version
\ifx\cleanversion\undefined
\else
 \renewcommand{\remark}[1]{\iffalse #1 \fi} % 删除remark的内容，remark 作为一个临时标记
 \renewcommand{\removed}[1]{\iffalse #1 \fi} 
 \renewcommand{\revision}[1]{#1}
\fi

\newcommand{\circled}[1]{#1}

\begin{document}

\title{Proteus: Shapeshifting Desktop Visualizations for Mobile via Multi-level Intelligent Adaptation}


\author{Can Liu}
  \email{can.liu@ntu.edu.sg}

\affiliation{%
  \institution{Nanyang Technological University}
  \city{Singapore}
  \country{Singapore}
}

\author{Sizhe Cheng}
\email{sizhe.cheng@ntu.edu.sg}
\affiliation{%
  \institution{Nanyang Technological University}
    \city{Singapore}
  \country{Singapore}
}

\author{Feng Liang}
\email{feng.liang@ntu.edu.sg}
\affiliation{%
  \institution{Nanyang Technological University}
    \city{Singapore}
  \country{Singapore}
}

\author{Zhibang Jiang}
\email{zhibang.jiang@gmail.com}
\affiliation{%
  \institution{Providence Health \& Services}
  \city{Renton}
  \state{Washington}
  \country{USA}
}

\author{Lingru Huang}
\email{huanglingru745@gmail.com}
\affiliation{%
  \institution{The Hong Kong Polytechnic University}
  \city{Hong Kong}
  \country{China}
}

\author{Kavinda Athapaththu}
\email{kavinda.athapaththu@ntu.edu.sg}
\affiliation{%
  \institution{Nanyang Technological University}
    \city{Singapore}
  \country{Singapore}
}

\author{Yong Wang}
\email{yong-wang@ntu.edu.sg}
\authornotemark[1]
\affiliation{%
\authornote{Y. Wang is the corresponding author.}
  \institution{Nanyang Technological University}
    \city{Singapore}
  \country{Singapore}
}

% \affiliation{%

% \affiliation{%

% \affiliation{%

\renewcommand{\shortauthors}{Trovato et al.}

\newcommand{\toolname}{\textit{Proteus}}

\begin{abstract}
With the rise of mobile-first consumption, users increasingly engage with data visualizations on mobile devices. However, the vast majority of existing visualizations are originally authored for desktop environments. Due to significant differences in viewport size and interaction paradigms, directly scaling desktop charts often results in illegible text, information loss, and interaction failures.
To bridge this gap, we propose an automated framework to adapt desktop-based visualizations for mobile screens. By systematically categorizing the operations involved in the adaptation process, we establish a multi-level design space. This space defines evolution rules spanning from the global topology level, through the reference frame level, down to the visual elements level. Guided by this theoretical framework, we developed \toolname{}, a large language model–driven multi-agent system that automatically parses the online visualizations, predicts optimal transformation strategies within the design space, and generates equivalent, highly readable visualizations for mobile devices.
Case studies and an in-depth user study with 12 participants demonstrate the effectiveness and usability of \toolname{}.

% With the rise of mobile-first consumption, users increasingly engage with data visualizations on mobile devices. However, the vast majority of existing visualizations are originally authored for desktop environments. Due to significant disparities in pixel density and interaction paradigms, direct scaling of desktop charts often results in illegible text, information loss, and interaction failure.
% To bridge this gap, we propose an automated workflow to adapt desktop-based visualizations for mobile screens.
% By systematically categorizing valid operations during the adaptation process, we establish a Multi-level Design Space. This space defines evolution rules spanning from the global topology level and reference frame level to the visual elements level, revealing how constraints propagate and interact across these hierarchical layers.
% Guided by this theoretical framework, we developed a Large Language Model-driven multi-agent framework, \toolname{}, which automatically parses the semantic content of source visualizations in HTML, predicts optimal transformation strategies within the design space, and generates equivalent, highly readable visualizations for mobile devices.
\end{abstract}





\received{20 February 2007}
\received[revised]{12 March 2009}
\received[accepted]{5 June 2009}

\maketitle

\begin{CJK}{UTF8}{gbsn}

\section{Introduction}

% The ubiquity of mobile computing has fundamentally reshaped information access. With the rise of mobile-first consumption, users increasingly engage with data visualizations on handheld devices, ranging from monitoring financial dashboards to tracking public health statistics. Despite this shift, the vast majority of visualizations are still authored in desktop environments, implicitly designed for landscape orientation, high pixel precision, and mouse-hover affordances.
% This disparity creates a critical ``Authoring-Consumption Gap."
% When visualizations designed for large monitors are ported to small mobile screens, the user experience frequently deteriorates. Simple scaling often results in illegible text, overlapping elements, and the loss of interactive precision.
% To address this, the visualization community has developed techniques for Responsive Visualization (RV). Seminal work by Hoffswell et al.~\cite{hoffswell2020techniques} established a design space for flexible visualizations, categorizing techniques into layout adjustments, scale modifications, and encoding changes.

The ubiquity of mobile computing has fundamentally reshaped information access. With the rise of mobile-first consumption, users increasingly engage with data visualizations on handheld devices, ranging from monitoring financial dashboards to tracking public health statistics. Despite this shift, the vast majority of visualizations are still authored in desktop environments, implicitly designed for landscape orientation, high pixel precision, and mouse-hover affordances.
This disparity creates a critical authoring–consumption gap.
When visualizations designed for wide monitors are ported to small mobile screens, the user experience frequently deteriorates.
Simple scaling often results in illegible text, overlapping elements, and the loss of interactive precision.
To address this, the visualization community has developed techniques for responsive visualization (RV). Seminal work by Hoffswell et al.~\cite{hoffswell2020techniques} established a design space for flexible visualizations, categorizing techniques into layout adjustments, scale modifications, and encoding changes.
\revision{More recent systems such as MobileVisFixer~\cite{wu_mobilevisfixer_2020} further automate parts of this process by operating on rendered SVG visualizations through an intermediate structured representation.
However, such systems remain tied to a predefined representation and transformation pipeline, which can limit the range of adaptations they support in practice.}
% More recent systems such as MobileVisFixer~\cite{wu_mobilevisfixer_2020} further automate parts of this process, but typically operate on structured visualization specifications (i.e., Vega-Lite~\cite{satyanarayan2017vegalite}) and therefore support only a limited subset of visualizations.
% \todo{Add Cicero}
\revision{Cicero~\cite{cicero2022} further advances this line of work by providing a declarative grammar for responsive visualization transformations, allowing design-agnostic adaptation rules to be reused across visualizations; nevertheless, like many prior systems, its scope remains centered on structured specifications rather than the diverse implementation styles found in practice.}
In practice, many web charts are implemented with heterogeneous tool-chains (such as D3, Plotly, or custom SVG/HTML), making it difficult for such specification-bound approaches to generalize across platforms and implementation styles.
Moreover, existing RV techniques predominantly rely on a flat taxonomy of geometric heuristics.
They treat adaptation as a layout puzzle, rearranging bounding boxes or shrinking elements to fit the view-port.
While effective for simple charts, this geometry-centric paradigm falls short when handling complex data. It lacks semantic understanding, often resorting to hiding labels or truncating text blindly, which compromises information fidelity. In addition, most methods treat operations in isolation and do not account for how a high-level decision (for example changing a chart type or reflowing a grid) structurally constrains lower-level elements (for example axis ticks, legends, and annotations).

To bridge this gap, we argue that effective mobile adaptation requires more than layout responsiveness; it demands automated semantic re-authoring.
In this paper, we present \toolname{}, an automated framework driven by Large Language Models (LLMs) that transforms desktop charts into mobile-optimized visualizations.
% Our approach is grounded in a novel multi-level design space, which systematizes adaptation not as a flat list of rules, but as a hierarchy of constraint propagation.
\revision{Our approach is grounded in a novel multi-level design space that organizes adaptation as a hierarchy of decisions and refinements, rather than a flat list of rules.}
Unlike prior work, our framework operates across three distinct layers:
\emph{Level 1: Global Topology}, which makes macroscopic decisions (for example axis transposition and grid reflow) that define the overall spatial structure of the visualization;
\emph{Level 2: Reference Frame}, which adjusts axes, legends, and scales (for example tick decimation and scroll injection) to balance quantitative precision with legibility; and
\emph{Level 3: Visual Elements}, which performs micro-level refinements (for example semantic abbreviation and label externalization) on marks and text to optimize the use of data ink on small screens.

Guided by this hierarchical framework, \toolname{} employs a multi-agent architecture.
By leveraging the semantic reasoning of LLMs, the system parses the source visualization to predict optimal transformation strategies, and generates executable code.
This allows \toolname{} to perform sophisticated adaptations, such as converting a static small-multiple view into an interactive carousel or semantically shortening categorical labels.
We present case studies to show how \toolname{} works for different types of visualizations from different .
Also, we evaluate \toolname{} in a comparative user study with 12 participants against a multi-agent LLM-based baseline on a diverse benchmark of real-world web visualizations.
The results show that our approach achieves higher render success rates and significantly better ratings across five dimensions: execution completeness, data fidelity, perceptual readability, visual aesthetics, interaction reasonableness, according to Wilcoxon signed-rank tests ($p < 0.05$ for fidelity/readability and $p < 0.001$ for aesthetics and interaction).

In summary, our contributions are as follows:
\begin{itemize}
    \item A multi-level design space that advances beyond flat responsive taxonomies by modeling the hierarchical propagation of constraints from global structure to individual visual marks.
    \item Proteus, an LLM-driven multi-agent framework that automates semantic parsing, planning, code generation, and quality evaluation for mobile visualization adaptation.
    \item Case studies and a controlled user study with 12 participants demonstrating that our semantic re-authoring approach significantly outperforms a strong LLM-based baseline across rendering success and all five evaluation dimensions on mobile devices.
\end{itemize}

% The online gallery and additional details about the system are available at 
More example visualizations processed by \toolname{} for mobile devices are available at
\url{https://vis2mobile.vercel.app}.

\section{Related Work}

We situate our work at the intersection of responsive visualization techniques, automated design systems, and the emerging application of multi-agent systems.

\subsection{Responsive Visualization Techniques}

Adapting visualizations for mobile devices involves balancing the reduction of screen space with the preservation of information density and readability~\cite{hoffswell2020techniques}. Early approaches relied on interactive exploration techniques, such as zooming, panning, and fisheye distortions, to navigate large charts on small screens~\cite{bederson2000fisheye}.
% Later, researchers formalized specific adaptation actions, such as resizing, repositioning, and modifying encodings, leading to declarative libraries like RespVis~\cite{andrews_respvis_2023}, which allow authors to manually define breakpoints and rules.
\revision{Later, researchers formalized specific adaptation actions, such as resizing, repositioning, and modifying encodings, leading to declarative libraries like RespVis~\cite{andrews_respvis_2023} and Cicero~\cite{cicero2022}, which allow authors to manually define rules, and reusable transformation specifications.}
To reduce the manual effort of defining rules, recent works have shifted towards automated layout optimization. MobileVisFixer~\cite{wu_learning_2021} employs a reinforcement learning framework trained on human preferences to automatically re-scale and reposition SVG elements.
Similarly, Zeng et al.~\cite{zeng_semiautomatic_2024} utilized simulated annealing to preserve spatial relationships in dashboards, while constraint-based methods trigger adaptations based on element legibility rather than static screen widths~\cite{schottler_constraint-based}.

% \revision{These approaches are effective when adaptation can be formulated as optimization over a predefined layout space, but they generally require task-specific objectives, hand-designed constraints, or limited transformation vocabularies. In contrast, our work frames mobile adaptation as a generative process over a multi-level design space. The visual specification in our framework serves as an intermediate representation that makes the generated design decisions explicit and executable, rather than encoding a fixed set of manually authored transformation rules.}
These existing frameworks primarily operate at a flat level.
They treat adaptation as a bounding-box rearrangement problem and cannot perform content-based re-authoring, such as linguistically abbreviating labels, or switching chart types to suit the data's message on a smaller canvas. This limitation often results in designs that are technically responsive but cognitively cluttered.

\subsection{Automated Visualization Systems}
Automated visualization design aims to recommend or generate effective charts, given data and tasks, reducing the effort of manual authoring—an effort that remains challenging even for experts~\cite{qin_making_2020}. Over the years, prior work has evolved along three major directions.
First, rule- and constraint-based systems make design knowledge explicit and computable, from foundations such as APT~\cite{mackinlay1986APT} to later recommender frameworks (e.g., ShowMe~\cite{showme07}, Voyager~\cite{voyager15}, and Draco~\cite{draco18}), as well as query-driven exploratory approaches~\cite{seedb14,foresight17}.
Second, data-driven methods learn charting decisions from large corpora of data-visualization pairs~\cite{vizml18,data2vis18,luo2021nl2vis,liu2021advisor,nl4dv20}. Third, visualization reuse and recovery techniques infer structure from existing artifacts, e.g., recovering encodings from rendered charts or extracting reusable templates to be repopulated with new data, enabling authoring by reuse rather than from scratch~\cite{revision11,poco17,chart-rcnn23,liu2020autocaption,mi3-21,datawink25, lai2020automatic, simvecvis}.

Recently, LLMs have encouraged agentic workflows that decompose visualization generation into planning, coding, and validation, sometimes with multimodal feedback to improve correctness\cite{ouyang_nvagent_2025,goswami_plotgen_2025,wolter_multiagent_2025}.
However, most such systems focus on Text-to-Vis generation, while the Vis-to-Vis adaptation setting remains underexplored: given a desktop visualization, the system must infer its structure and interaction intent and then re-author it for mobile constraints. Prior mobile-oriented work (e.g., MobileVisFixer\cite{wu_mobilevisfixer_2020}) primarily targets layout repair via learning-based methods, leaving adaptation of multi-view organization and interaction redesign largely open.

\subsection{Multi-Agent Systems for Visualization}

Multi-agent and mixed-initiative ideas have long been explored to make complex interactive tasks more manageable—systems distribute reasoning and responsibilities across cooperating components while keeping humans in control through direct manipulation and incremental refinement~\cite{horvitz99, mas-survey09}.
In visualization, collaborative platforms introduced early coordination mechanisms such as shared annotations and versioning~\cite{manyeyes07, senseus07}.
Building on classic foundations in distributed problem solving and collective intelligence~\cite{dps-book88, wooldridge-book09, collective-intelligence-book10}, recent advances in large language models (LLMs) have further encouraged multi-component architectures where role-specialized agents or modules can plan, invoke tools, reflect, and coordinate model–tool pipelines~\cite{autogpt23, camel23, reflexion23, voyager_agent23, hugginggpt23, ijcai2024p890}.
However—as visualization design automation remains challenging—prior systems mostly rely on heuristics or fixed templates for chart generation or scaling across devices rather than truly agent-based coordination~\cite{heer2012interactive, endert2014semantics, amershi2019guidelines, agdebugger25}.
Athanor~\cite{liu2026athanor} instead uses multiple collaborating agents to enable more accurate intent interpretation for adding interaction to an existing visualization.
Our work adopts MAS-inspired principles to structure collaborative LLM agents for Vis-to-Vis adaptation, enabling semantic reasoning over desktop visualizations and coordinated decisions to produce faithful yet optimized mobile counterparts.

\section{A Multi-level Design Space for Mobile Visualization Adaptation}
\label{sec:design_space}

The transition from desktop to mobile visualization is not a trivial graphical rescaling task; it is a reconstruction process governed by competing constraints. Desktop environments are characterized by landscape orientation, high pixel precision, mouse-hover affordances, and expansive screen real estate. In contrast, mobile environments impose strict physical and interaction limitations: portrait orientation, fat-finger input, and narrow horizontal bandwidth. To navigate this trade-off systematically, we first identify the core design requirements and then map them to a hierarchical design space.

\subsection{Design Requirements for Mobile Visualization Adaptation}

To bridge the gap between desktop authoring and mobile consumption, we identify four design requirements. These are derived from the fundamental conflict between \textit{information density} (high on desktop) and \textit{perceptual bandwidth} (low on mobile).

\begin{itemize}

\begin{figure*}
    \centering
    \includegraphics[width=\textwidth]{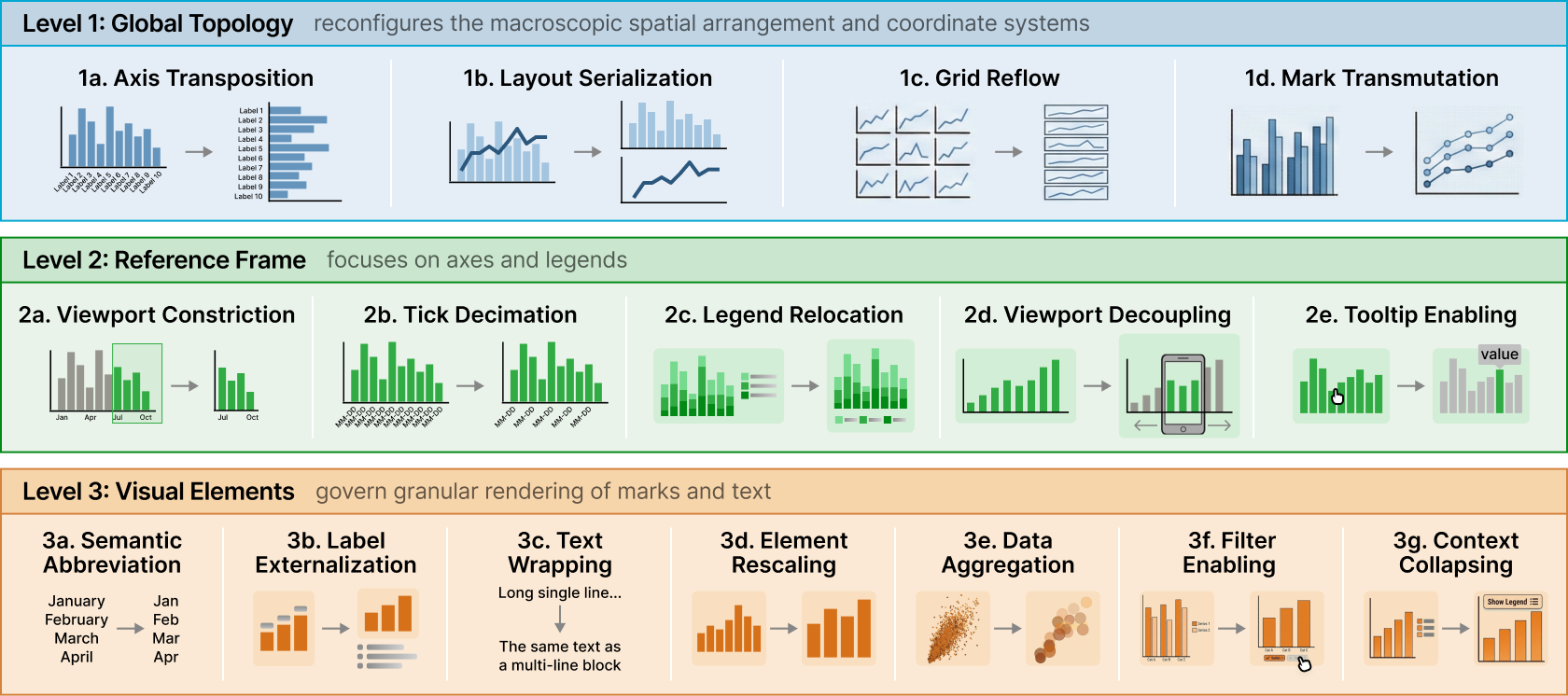}
    \caption{The proposed multi-level design space for mobile visualization adaptation. 
    It organizes transformation operators into three levels: (1) \emph{Global Topology}, which restructures the overall spatial layout and coordinate systems; 
    (2) \emph{Reference Frame}, which adjusts axes, ticks, legends, and viewports; and 
    (3) \emph{Visual Elements}, which refines the rendering of marks and text for mobile screens.}
    \label{fig:design_space}
    \Description{A conceptual diagram illustrating a hierarchical design space for adapting visualizations to mobile devices. The framework is organized into three layers, covering overall layout restructuring, adjustments to coordinate references such as axes and legends, and fine-grained refinement of graphical marks and text.}
\end{figure*}

\item \textbf{R1. Ensure Perceptual Scalability.}
Mobile screens suffer from a scarcity of pixels.
A direct reduction in size leads to occlusion and illegibility.
This requires aggressive operations (such as decimation or re-layout) to ensure that all rendered elements (text, axes, marks) remain legible and distinguishable, prioritizing the visibility of essential signals over decorative fidelity.

\item \textbf{R2. Preserve Semantic Equivalence.} 
\revision{Simply resizing a visualization can hide information that is necessary for interpretation. For example, hiding axis labels may make a chart difficult or impossible to understand. Therefore, adaptation should go beyond layout adjustment and selectively revise visual elements while preserving the original meaning of the data. To do so, the system needs to consider the data context and the role of each visual component.}

\item \textbf{R3. Exchange Space for Time.} 
Mobile devices lack the spatial width of desktops but possess superior affordances for touch and scroll navigation. The framework can leverage this by converting \textit{spatial density} into \textit{temporal interaction}. Static, crowded elements (like small multiples or dense axes) should be transformed into dynamic, scrollable, or collapsible components, allowing users to explore high-resolution data through gesture-based navigation rather than visual scanning.

\item \textbf{R4. Respect Cross-Level Dependencies.} 
A visualization is a hierarchical scene graph. A change in the global chart type (e.g., transposing axes) alters the constraints for lower-level elements (e.g., label orientation). The design space should formally model these dependencies. Rather than a set of isolated tweaks, adaptation should be modeled as a top-down propagation of constraints where topological decisions dictate the valid parameter space for atomic elements.

\end{itemize}

The requirements above necessitate a framework that handles adaptation at different levels of abstraction. However, a flat list of operators is insufficient for the hierarchical nature of \textbf{R4}. To structure this space systematically, we decompose the visualization into three levels.
We observe that a visualization is constructed in three distinct logical stages: defining the spatial substrate, establishing the reference context, and populating the data content. Accordingly, we partition our design space into three layers as illustrated in \autoref{fig:design_space}:

\begin{itemize}
\item \textbf{Level 1: Global Topology.} 
This level dictates the macroscopic arrangement, encompassing container layout, coordinate systems, and facet topology. 
It defines \textit{where} content can exist and determines the bounding boxes for all lower levels. 
In mobile adaptation, this often necessitates \textbf{structural reconfiguration} (addressing the density and hierarchy requirements in R1 and R4), such as transforming composite, high-density desktop dashboards into decomposed or serialized layouts (e.g., splitting a multi-view dashboard into separate mobile pages or reflowing a grid of charts into a vertical stack).

\item \textbf{Level 2: Reference Frame.} 
% This level serves as the semiotic bridge between data values and screen coordinates, managing axes, legends, and scales. 
\revision{This level manages the visual elements that help users interpret data values in the allocated space, including axes, legends, and scales.}
It defines \textit{how} to measure and interpret the space allocated by Level 1. 
Adaptation here focuses on contextual optimization (in line with R1 and R2): axes and legends must be resized or repositioned to maximize space utilization, and in some cases auxiliary guides or legends must be explicitly added to preserve semantic interpretability under mobile constraints.

\item \textbf{Level 3: Visual Elements.} 
This level governs the visual marks (bars, points, lines) and the text elements, that is, the actual content that populates the reference frame. 
Since mobile screens often render desktop-sized elements spatially prohibitive (e.g., long labels causing overflow), this layer handles granular adjustments (e.g., text wrapping, semantic abbreviation, and mark rescaling) to ensure perceptual scalability and semantic fidelity as required by R1 and R2, while also preparing elements to participate in interaction patterns that realize R3.
\end{itemize}

\subsection{Level 1: Global Topology}
\label{sec:level1}
The highest level addresses the macroscopic structure of the visualization. Adaptation here involves altering the spatial arrangement of views, defining the container boundaries, and fundamentally re-orienting the coordinate system. Since these decisions determine the bounding boxes and spatial logic of the reference frames and visual elements, they must be resolved first.

\subsubsection{Coordinate Transformation}
One of the most salient disparities between desktop and mobile environments is the inversion of the aspect ratio (typically wide landscape vs. narrow portrait).
Coordinate transformation operators address this by remapping the spatial encoding channels.

\begin{itemize}
\item \textbf{Axis Transposition.}
Mobile devices typically favor a portrait orientation (height > width), which stands in sharp contrast to the landscape layouts common on desktop displays.
To address this geometric inversion, transposing the axes becomes a fundamental adaptation strategy.
For ordinal or nominal data (e.g., bar charts), limited horizontal bandwidth often constrains the number of categories that can be displayed and can lead to overly dense labels along the horizontal axis.
This operator swaps the horizontal and vertical axes, mapping the categorical dimension onto the y‑axis.
Such transposition leverages the native vertical scrolling affordance of mobile interfaces, allowing more data points to be shown without compression.
Moreover, because most text (e.g., English) is laid out horizontally, placing category labels along the vertical axis also helps avoid the crowding issues that would otherwise arise from densely packed, rotated labels.

\item \textbf{Layout Serialization.} 
On desktop displays, ample pixel bandwidth often allows multiple data layers to be overlaid within a single shared coordinate space for direct comparison. On mobile screens, however, such multilayer overlays can become difficult to read due to limited resolution and screen size. 
For layers that do not strictly need to be superimposed, this operator applies decomposition: it splits the combined view into two or more separate charts and serializes them in the layout.
This reduces visual clutter while preserving the essential comparative capability through sequential inspection rather than simultaneous overlay.

\end{itemize}

\subsubsection{Facet Management}
Small multiples (faceted views) present a significant challenge.
For example, a $3 \times 2$ grid effective on a monitor becomes illegible on a phone.

\begin{itemize}

\item \textbf{Grid Reflow}
This operator addresses the legibility collapse of spatially distributed small multiples on narrow screens. 
It modifies the layout topology by reducing the grid's column cardinality, effectively enforcing a serialization of the view. 
For instance, a compact $3 \times 2$ desktop grid is reflowed into a vertical stack ($N=1$), allowing each individual chart to expand to the full width of the device viewport.

\end{itemize}

\subsubsection{Encoding Transformation}
In certain scenarios, the geometric footprint of a desktop chart is fundamentally incompatible with mobile constraints, regardless of layout adjustments. This necessitates a transformation of the visual encoding itself to reduce horizontal spatial demand.

\begin{itemize}
\item \textbf{Mark Transmutation:}
Complex discrete geometries, such as grouped bar charts, suffer from severe crowding on narrow screens due to the high number of bars per tick. This operator transmutes the mark type into a more compact form.
For instance, converting grouped bars into multi-line charts maintains the multi-series comparison while reducing the required pixels per data point.

\end{itemize}

\subsection{Level 2: Reference Frame}
\label{sec:level2}

Once the global topology is established, the design space focuses on the reference frame. The axes and legends act as the bridge between raw data values and screen coordinates. Mobile adaptation here primarily concerns the trade-off between \textit{precision} (showing all ticks) and \textit{legibility} (showing readable ticks).

\subsubsection{Domain and Range Manipulation}
In mobile environments, where the available "pixel budget" is severely restricted, the default mapping strategies used on desktops often result in compressed or indistinguishable visual patterns. Adaptation at this level focuses on optimizing this projection to maximize discernibility.

\begin{itemize}
\item \textbf{Viewport Constriction:}
This operator addresses the loss of resolution caused by scaling down a large dataset to a small screen. Rather than preserving the global extent (0-100\%), it semantically crops the visible axis domain to a specific region of interest.
For example, a chart might zoom in on a dense cluster of values or default to showing only the most recent time window, forcing the viewer to scroll or interact to see the historical context.
This zooming trades global context for local fidelity.
Crucially, this constriction is typically paired with interactive affordances, such as a range slider, a mini-map brush, or pan gestures, enabling users to dynamically traverse the hidden data.
This approach implements a "Focus+Context" strategy, trading immediate global visibility for local fidelity while preserving navigational access.

\end{itemize}

\subsubsection{Axes Optimization Strategy}
On small or horizontally constrained displays, axes are often the first component to become problematic: as tick labels compete for limited width, they begin to overlap, quickly rendering the scale unreadable. This family of operators focuses on preserving the interpretability of axes under compression by either reducing the number of visible ticks when interpolation is possible, or reorienting labels when every category must remain explicitly shown.

\begin{itemize}
    \item \textbf{Tick Decimation:}
    For continuous attributes, such as quantitative or temporal scales, reducing label density rarely compromises information retrieval due to the user's capacity for linear interpolation. For instance, reducing a 12-month axis to quarterly ticks (``Jan", ``Apr", ``Jul", ``Oct") allows users to instinctively infer the positions of intermediate months (e.g., ``Feb", ``Mar"). This operator improves legibility by enforcing sparsity while relying on the predictable nature of the scale to preserve the user's mental model.

    \item \textbf{Label Rotation:} 
    For nominal data attributes (e.g., specific category names), tick decimation is often inappropriate because users cannot infer missing values via interpolation, unlike with continuous attributes.
    When preserving the visibility of \textit{every} category is mandatory but horizontal space is insufficient, this operator acts as a critical adaptation strategy. By angling text (typically $45^\circ$ or $90^\circ$), it utilizes vertical space to prevent overlap.
    Importantly, the effectiveness of label rotation is language-dependent: certain writing systems (e.g., Mongolian script) are inherently designed for vertical layout, while some (e.g., Chinese characters) support both horizontal and vertical orientations with relatively minor legibility loss.
\end{itemize}

\subsubsection{Legend Adaptation}
Legends often consume valuable horizontal space on desktop displays. Adjusting the position of legends usually does not incur any loss of information.

\begin{itemize}
    \item \textbf{Repositioning:} Changing the placement of legends typically does not lead to information loss. For example, legends can be moved from a right-hand sidebar to the top or bottom of the chart, or converted into inline labels that are attached directly to the corresponding visual elements (e.g., lines in a line chart).
    In addition, legends can often be turned into interactive buttons, allowing users to tap them to customize which data series are shown.
\end{itemize}

\subsubsection{Interaction Injection}
Where static spatial resolution fails, temporal interaction compensates. We introduce two operators that shift the burden from visual perception to manual manipulation, adhering to the principle of \textit{Details-on-Demand}~\cite{shneiderman1996eyes}.

\begin{itemize}

\item \textbf{Viewport Decoupling:} %  (\texttt{setPanInteraction})
When data density exceeds the physical pixel limit, compressing the entire dataset renders it illegible.
This operator decouples the logical chart width from the physical viewport width. It enables navigation through diverse mechanisms: direct manipulation (panning the x-axis or scrolling the virtual canvas) or proxy control (using an auxiliary widget like a range slider).
This maintains a high-resolution data representation, allowing users to explore the continuum sequentially rather than simultaneously.

\item \textbf{Tooltip Enabling:} 
As a consequence of aggressive tick decimation (Level 2) and mark rescaling (Level 3), the precise values of data points are often visually abstracted in the static view. This operator injects an interactive layer, typically triggered by touch or a long-press, to display exact values via overlays. This compensates for the loss of static fidelity, ensuring that quantitative precision is retrievable despite the simplified visual presentation.

\end{itemize}

\subsection{Level 3: Visual Elements}
\label{sec:level3}

The deepest level of the hierarchy governs the ``atomic'' constituents of the visualization: the geometric marks representing data points and their associated textual annotations. While upper levels define the container and reference frame, this level manages the actual \textit{data ink}. Constraints from the global-topology level (layout) and reference-frame level (range) propagate downward, often creating extreme spatial scarcity for these fundamental elements. Adaptation here focuses on maintaining \textbf{Semantic Fidelity (R2)} and \textbf{Interaction Feasibility (R3)} at the granular level.

\subsubsection{Semantic Text Adaptation}
Text is the most rigid element in visualization design; unlike geometry (bars, lines, or areas), it cannot be linearly scaled down without losing legibility. Consequently, text adaptation on mobile devices requires operations that prioritize \textit{meaning} over \textit{form}.

\begin{itemize}
\item \textbf{Semantic Abbreviation:}
This operator addresses the conflict between limited horizontal bandwidth and long nominal labels. Instead of naive truncation (which destroys meaning, e.g., ``South...'' could be ``South Dakota'' or ``South Carolina''), this operator employs domain-specific logic to compress string length while preserving semantic distinguishability. For instance, ``United States'' is mapped to ``USA'', and ``January 2023'' is compressed to ``23 Jan''.

\item \textbf{Label Externalization:} 
When data density prohibits in-situ labeling (placing labels directly next to marks), text occlusion becomes more severe in mobile environments. This operator decouples the label from its spatial position by replacing the on-chart text with a compact indicator. The full textual content is then relocated to an external list or legend below the chart, resolving the clutter without discarding information.

\item \textbf{Text Wrapping:} 
Long titles or annotations often exceed the viewport width in portrait mode. This operator utilizes vertical space to compensate for horizontal scarcity by introducing line breaks. It transforms a single-line string into a multi-line block, ensuring the full context is visible without forcing horizontal scrolling.
\end{itemize}

\subsubsection{Geometry and Data Reduction}
On small screens, the physical size of visual marks must balance two competing objectives: they must be small enough to prevent overplotting, yet large enough to be accurately perceived and touched.

\begin{itemize}
\item \textbf{Element Rescaling:}
Although the screen dimensions contract, visual marks cannot undergo a simple linear down-scaling. Such a reduction would render elements too minute for accurate interaction. Instead, this operator enforces a relative enlargement of geometric primitives (e.g., points or bars), ensuring that they maintain a sufficient physical size to accommodate coarse touch inputs, even if this disproportionate scaling necessitates a reduction in data density.

\item \textbf{Data Sampling and Aggregation (\texttt{sampleData}):} 
When the cardinality of the dataset exceeds the available screen pixels, rendering every data point results in visual noise and rendering bottlenecks. This operator reduces visual density by either selecting a representative subset (sampling) or combining proximal data points into summary statistics (i.e., aggregation or binning), ensuring that the overarching trend remains discernible.
\end{itemize}

\subsubsection{Visibility and Progressive Disclosure}
Mobile interfaces necessitate a strategy of progressive disclosure, showing only what is essential and hiding secondary information until requested.

\begin{itemize}
\item \textbf{Filter Enabling:}  
For charts with high-cardinality elements, such as grouped bar charts, displaying all series simultaneously on a narrow screen often results in overly thin, hard-to-read bars. This operator mitigates overcrowding by introducing external controls, such as segmented buttons or filter chips. Instead of presenting a dense visualization, it allows users to manually toggle the visibility of specific data subsets. For example, a grouped bar chart showing trends across multiple categories can be transformed into a cleaner, more legible view where users can focus on one series at a time through a button strip, trading immediate cross-series comparison for enhanced visual clarity.

\item \textbf{Context Collapsing:} 
Auxiliary context, such as lengthy subtitles or secondary legends, consumes valuable data ink.
This operator encapsulates such auxiliary information in interactive toggles that users can expand on demand. In this way, static details are hidden by default to maximize the plot area, yet remain easily accessible when needed. This pattern is also common on mobile devices, where limited screen space encourages collapsing secondary details behind tappable controls.
\end{itemize}

\subsection{Constraint Propagation and Inter-level Dynamics}

Crucially, the three levels in our design space do not operate in isolation. Decisions made at higher levels impose hard constraints on what is feasible at lower levels, and lower-level requirements can in turn trigger structural adjustments upstream.

Changes at the global topology level propagate downward. Operations such as axis transposition, layout serialization, or grid reflow alter the aspect ratio and container boundaries, which in turn constrain how legends can be repositioned and whether labels must be shortened or externalized (\autoref{sec:level2}, \autoref{sec:level3}). For example, when a dense small-multiple grid is reflowed into a vertical stack, legends may need to move to the top or be converted into inline labels, and long category names may require semantic abbreviation or label externalization to remain readable.
At the reference frame level, reducing tick density (tick decimation) is often paired with interactive mechanisms such as tooltips or filters, so that users can still retrieve precise values on demand despite fewer visible labels (\autoref{sec:level2}). Likewise, when the visible range is narrowed through viewport constriction, it is commonly combined with viewport decoupling, allowing users to access the full data extent via panning or range selection.
At the visual elements level, operators such as element rescaling are directly determined by constraints from upper levels: once the viewport has been constricted and the axis range fixed, mark sizes must be chosen so that they remain both perceptible and touchable within the available pixels. In other words, fine-grained changes to marks and text are not arbitrary tweaks, but must respect the layout and reference-frame decisions already in place.
These inter-level dynamics help explain why simple rule-based heuristics that treat operations independently often fail on mobile: improving one aspect (for example, density via tick reduction) without coordinating compensating mechanisms (such as tooltips, scrolling, or context collapsing) can easily break readability or fidelity.

\section{Proteus}

% \section{The Proteus Framework}
\label{sec:system_framework}

\begin{figure*}[ht]
    \centering
    \includegraphics[width=\textwidth]{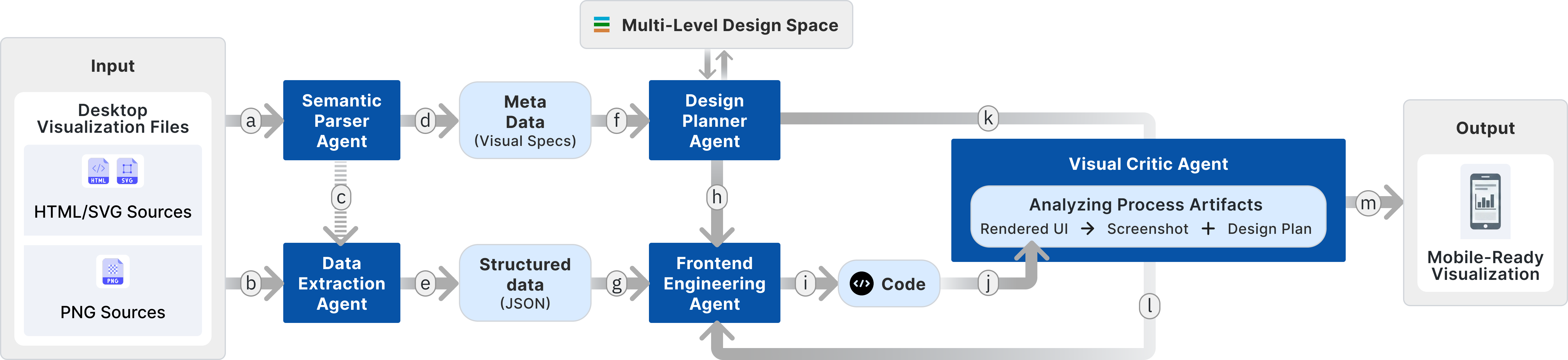}
    \caption{The multi-agent framework of \toolname{} for desktop-to-mobile visualization adaptation.}
    \Description{An overview of the Proteus system architecture, depicting multiple coordinated agents that collaboratively convert desktop visualizations into mobile-friendly versions through a staged process.}
    \label{fig:pipeline}
\end{figure*}

Guided by the Multi-level Design Space defined in \autoref{sec:design_space}, we developed \toolname{}, a Large Language Model (LLM) driven multi-agent system. \toolname{} automates the transformation of visualizations, moving from a desktop visualization or source code to a fully responsive mobile component.

\subsection{Agent Roles and Responsibilities}

To replicate the workflow of a human visualization expert, \toolname{} is architected as a collaborative team of five specialized agents (as illustrated in \autoref{fig:pipeline}): the \textit{Semantic Parser}, the \textit{Data Extractor}, the \textit{Design Planner}, the \textit{Frontend Engineer}, and the \textit{Visual Critic}. These agents operate within a closed-loop iterative environment, ensuring that the final output not only compiles but also strictly adheres to mobile readability constraints.

\begin{itemize}
\item \textbf{Semantic parser agent.}
The pipeline initiates with the semantic parser. Upon ingesting the desktop source code (HTML/SVG), this agent first orchestrates a rendering process to capture a high-fidelity raster snapshot (PNG) of the visualization. Crucially, it executes a multimodal analysis, correlating the explicit \textit{vector structure} (DOM/SVG hierarchy) with the perceptual bitmap layout. By bridging code-level semantics with pixel-level spatial relationships, its primary objective is to perform robust Visualization Deconstruction, identifying the global chart topology (e.g., Grouped Bar Chart) and segmenting logical components (axes, legends, marks) before passing control to the extraction layer.

\begin{lstlisting}[style=beautiful_listing]
You need to act as a **Semantic Parser Agent** in a visualization-to-mobile transformation pipeline. 

Your primary objective is to execute a multimodal analysis, correlating the explicit vector structure (DOM/SVG hierarchy) with the perceptual bitmap layout. You will perform robust Visualization Deconstruction and extract real data.
## Project Structure & Context
- `original_visualization`: 
    - `desktop.{{html, svg}}`: the original source code of the visualization tailored for desktop.
    - `desktop.png`: the rendered original visualization tailored for desktop.
    - `desktop_on_mobile.png`: the original visualization directly rendered in a mobile aspect ratio.
### Source Code
Below is the source code of the original SVG or HTML file: {source_code}
### Visual Renderings
The first image is `desktop.png`. The second image is `desktop_on_mobile.png`.
[IMAGE_PLACEHOLDER: desktop.png (tailored for desktop)]
[IMAGE_PLACEHOLDER: desktop_on_mobile.png (rendered in mobile aspect ratio)]
## Task and Requirements
1. **Multimodal Analysis & Correlation**: Analyze the source code alongside the provided images. Bridge the code-level semantics with pixel-level spatial relationships.
2. **Visualization Deconstruction**: 
    - Identify the global chart topology (e.g., Grouped Bar Chart, Line Chart, Scatter Plot, etc.).
    - Segment and list all logical components (axes, legends, marks, titles, annotations, grid lines).
3. **Mobile Rendering Analysis**: Analyze the `desktop_on_mobile.png` image. Explicitly document the visual issues and constraints when this desktop layout is forced into a mobile aspect ratio (e.g., overlapping text, squished marks, hidden elements).
4. **Codify Data (Crucial)**: Extract and structure the **real data** from the source SVG (coordinates, colors, labels, raw data values) or HTML files. **No fake data allowed.** Map the visual properties back to the underlying data points.
## Output Format
Write your output in Markdown format. It must include:
1. **Topology & Component Segmentation**: A detailed breakdown of the chart type and its logical components.
2. **Mobile Constraints Analysis**: Identified issues in the mobile aspect ratio render.
3. **Extracted Data**: A cleanly structured JSON or tabular format of the exact data extracted from the code, ready to be used by the downstream planner and engineering agents.
\end{lstlisting}

\item \textbf{Data extraction agent.}
Employing a customized reverse engineering approach, this agent utilizes the parsed vector structure to recover the underlying dataset and visual specifications (e.g., scale domains, color mappings, and axis ranges). Unlike generic tools or simple pixel-based resizing, it executes case-specific logic to extract precise geometric attributes (e.g., height, position, and color) from the source code. By integrating these visual cues with metadata-derived mapping rules, or by directly parsing explicit text labels where data is visible, the agent computes the exact numerical content.
This reconstructed data is consolidated into a structured JSON format, ensuring that semantic fidelity is strictly preserved during the migration from the desktop source to the mobile implementation.

\begin{lstlisting}[style=beautiful_listing]
You are a highly specialized Data Extraction Agent.
Your task is to recover the underlying dataset and visualization specifications from given source code (e.g., SVG, Canvas, or chart-rendering scripts), and output the extracted data in a structured JSON format.
### Capabilities
- You may write and execute Python or JavaScript code to assist with parsing and extraction.
- You are allowed to perform reverse engineering on the source code to recover data.
### Instructions
1. Reverse Engineering
   - Analyze the provided source code and parsed vector structure.
   - Identify visual elements such as shapes, paths, bars, lines, and text.
2. Extract Visual Specifications
   - Recover key visualization parameters, including:
     - Scale domains
     - Axis ranges
     - Color mappings
     - Coordinate systems
3. Geometric Attribute Extraction
   - Extract precise geometric attributes such as:
     - Position (x, y)
     - Size (width, height, radius)
     - Color values
   - Do NOT rely on pixel-level heuristics; instead, use structural and code-level information.
4. Data Reconstruction
   - Map geometric attributes back to data values using:
     - Scale transformations
     - Axis mappings
     - Metadata-derived rules (if available)
   - If explicit text labels are present, directly parse and use them.
\end{lstlisting}

\item \textbf{Design planner agent.}
Armed with the extracted data and original visualization structure, the Design Planner acts as the system's architect. It references the design space (Section \ref{sec:design_space}) to generate a comprehensive transformation action plan. The agent adopts a hierarchical decision-making process: it first evaluates the disparity between desktop and mobile form factors to determine the necessity of global topology changes, such as axis transposition or layout serialization. Once the structural foundation is set, it decides on adjustments to the reference frame, including modifications to axes and legends, as well as the integration of complementary interactions. Finally, it specifies granular adaptations at the element level. This process culminates in a high-level blueprint that explicitly instructs the engineering agent on handling specific challenges, such as converting legends to inline labels or transforming the x-axis into a scrollable container, effectively mapping abstract constraints to concrete implementation steps.

% Armed with the extracted data and original visualization structure, the Design Planner acts as the system's architect. It references the design space (Section \ref{sec:design_space}) to generate a comprehensive transformation action plan. The agent adopts a hierarchical decision-making process: it first evaluates the disparity between desktop and mobile form factors to determine the necessity of global topology changes, such as axis transposition or layout serialization. Once the structural foundation is set, it decides on adjustments to the reference frame, including modifications to axes and legends, as well as the integration of complementary interactions. Finally, it specifies granular adaptations at the element level. This process culminates in a high-level blueprint that explicitly instructs the engineering agent on handling specific challenges, such as converting legends to inline labels or transforming the x-axis into a scrollable container, effectively mapping abstract constraints to concrete implementation steps.
% \todo{add prompt here}

\begin{lstlisting}[style=beautiful_listing]
You need to act as a **Design Planner Agent** to architect how to transform a static desktop visualization into an interactive, mobile-friendly component.

## Global Objectives
1. **Mobile-First UX**: Ensure touch-friendliness, readable typography, and responsive layouts.
2. **Premium Aesthetics**: Use modern UI principles (glassmorphism, vibrant palettes, smooth animations) to create a "wow" factor.
3. **Information Preservation**: Preserve as much information and intention presented in the desktop version as possible.

## Tech Stack Target
- Framework: Next.js 14 (App Router)
- Styling: TailwindCSS
- Visualization: Recharts
- Icons: Lucide React

## Inputs
### 1. Semantic Parser Output
Below is the deconstructed chart topology, mobile constraint analysis, and extracted data provided by the Semantic Parser Agent: {semantic_parser_output}
### 2. Vis2Mobile Design Action Space
Below is the document detailing the design space of visualizations and the action space for mobile transformation: {vis2mobile_design_action_space}
### 3. Visual References (Context)
[IMAGE_PLACEHOLDER: desktop.png (tailored for desktop)]
[IMAGE_PLACEHOLDER: desktop_on_mobile.png (rendered in mobile aspect ratio)]
## Task and Requirements
Armed with the extracted data, original visual intent, and the Design Action Space, you must generate a comprehensive transformation action plan. 
Adopt a **hierarchical decision-making process** in your plan:
1. **Global Topology Adjustments**: Evaluate the disparity between desktop and mobile form factors (referencing the Semantic Parser's constraint analysis). Determine if global topology changes are needed (e.g., axis transposition, layout serialization, changing chart types).
2. **Reference Frame Adjustments**: Decide on modifications to axes and legends (e.g., converting legends to inline labels, making the x-axis a scrollable container). Plan the integration of complementary mobile interactions (e.g., touch tooltips, swipe gestures).
3. **Element-Level Adaptations**: Specify granular adaptations such as mark sizing, typography adjustments, and color palette alignment with premium aesthetics.

## Output Constraints
- Just give a plan, no need to implement the code yet. 
- You must explicitly list **the actions to take that are in the Action Space and justify why you chose them**.
- If an action is not in the action space, state the reason and justify why existing actions are unsuitable.
- If some information must be omitted or moved to another view due to mobile constraints, mention it and write your justification.
- **Structure your response**: Start with a High-Level Strategy, followed by Detailed Implementational Steps mapping abstract constraints to concrete engineering instructions.
- Write your plan in pure Markdown format.
\end{lstlisting}

\item \textbf{Frontend engineering agent.}
To translate the extracted data and the Design Planner’s strategic scheme into a functional application, the Frontend Engineering agent operates within a specified mobile software framework. Working in a web-based environment (HTML, JavaScript, and CSS), the agent is provided with a pre-defined, empty mobile application scaffold and tasked with authoring TypeScript~\cite{typescript} files to implement specific visualization components and interaction logic.
The choice of TypeScript is deliberate; by merging TypeScript’s robust type system with JavaScript’s declarative syntax~\cite{react2013}, it enables the unified expression of data structures, interaction logic, and interface layouts within a single component. This cohesive format not only enhances code readability and maintainability but also empowers the agent to precisely map the design planner’s abstract constraints into a concrete, executable mobile implementation.

\begin{lstlisting}[style=beautiful_listing]
You should implement the mobile version of the visualization in the `src/components/Visualization.tsx` file.

## Tech Stack Target
- Framework: Next.js 14 (App Router)
- Styling: TailwindCSS
- Visualization: Recharts
- Icons: Lucide React

Carefully read the following files for instructions and more information:

- `transform-plan.md`: the detailed plan for transforming the visualization to a mobile version. YOU SHOULD FOLLOW THIS PLAN.
- `mobile-vis-design-action-space.md`: high-level description of the space for the actions that an agent can take to transform the visualization to a mobile version
- `original_visualization`:
  - `desktop.{html|svg}`: the original source code of the visualization that is tailored for desktop.
  - `desktop.png`: the rendered original visualization that is tailored for desktop.
  - `desktop_on_mobile.png`: the original visualization that is directly rendered in mobile aspect ratio.
- `src/components/Visualization.tsx`: The main component that renders the transformed mobile visualization.
\end{lstlisting}

\item \textbf{Visual Critic Agent.}
Our framework includes a visual critic agent, which performs a comprehensive multi-dimensional evaluation. Once the code is generated, the system renders the component and captures a screenshot (mimicking a real mobile device viewport). Utilizing vision-capable LLMs, the visual critic agent assesses the visualization against four critical criteria:
\begin{itemize}
\item \textbf{Data Fidelity.} Verifying that the rendered visual elements accurately reflect the extracted numerical data without distortion or loss.
\item \textbf{Plan Adherence.} Ensuring the implementation strictly follows the topological and structural directives issued by the \textit{Design Planner}.
\item \textbf{Text Readability.} Detecting legibility degradation caused by spatial compression. As highlighted in prior research~\cite{wu_mobilevisfixer_2020}, maintaining text readability is often the most significant challenge when migrating visualizations to constrained mobile environments. The agent rigorously scans for symptoms of poor adaptation, such as aggressive font downsizing that renders text indecipherable, label occlusion due to horizontal crowding, or insufficient contrast against complex backgrounds, ensuring textual information remains accessible despite the reduced screen real estate.

\item \textbf{Aesthetics.} Evaluating the overall visual appeal by balancing stylistic fidelity to the desktop source with mobile-native harmony. The agent assesses whether the adaptation preserves the original visual identity (e.g., color themes and stylistic tone) while ensuring the layout remains spatially coherent, well-spaced, and aesthetically pleasing within the compact mobile viewport.
\end{itemize}
Upon detecting deficiencies, the agent dynamically routes feedback based on the severity of the issue.
Implementation-level defects (e.g., styling inconsistencies or minor layout shifts) are addressed to the frontend engineering agent for immediate code refinement, whereas fundamental structural flaws or strategy mismatches trigger a regression to the design planner agent for architectural revision.
\end{itemize}

\begin{lstlisting}[style=beautiful_listing]
You are a Visual Critic Agent for mobile visualization. Your task is to evaluate the quality of a rendered mobile visualization and provide actionable feedback.
### Instructions
1. Rendering Inspection
   - Run `uv run python screenshot.py` to capture the mobile visualization as `mobile-version.png`.
   - Inspect the screenshot to detect rendering issues or runtime errors.
   - Optionally, use Playwright to interact with and verify the rendered result.
2. Evaluation Criteria
   Evaluate the visualization along the following dimensions:
   (a) Data Fidelity: Check whether visual elements accurately reflect the underlying data. Ensure no distortion, mismatch, or missing data.
   (b) Plan Adherence: Verify that the implementation follows the design plan (layout, structure, chart type).
   (c) Text Readability: Detect issues such as: Font sizes too small; Overlapping or occluded labels; Poor contrast; Ensure text remains legible in a mobile viewport.
   (d) Aesthetics: Evaluate visual quality, including: Layout balance and spacing, consistency with original style (e.g., colors), and verall visual clarity and harmony.

3. Issue Classification & Routing
   - If issues are minor (e.g., styling, spacing, small layout shifts):
     → Provide fixes for the frontend engineering agent.
   - If issues are structural (e.g., wrong chart type, poor layout strategy):
     → Escalate to the design planner for revision.
4. Output
   - List detected issues grouped by category.
   - Provide clear, actionable suggestions.
   - Indicate routing decision: [Frontend Fix] or [Planner Revision].
\end{lstlisting}

\subsection{Workflow Execution}
\toolname{} executes visualization adaptation as an iterative, artifact-driven workflow (\autoref{fig:pipeline}). Rather than treating mobile adaptation as a one-shot “resize” problem, the system progressively translates a desktop visualization into a mobile component through a sequence of intermediate artifacts that can be inspected, revised, and validated.

\textbf{Inputs and runtime.}
The pipeline accepts desktop visualization assets, including HTML/SVG source code and the corresponding rendered raster snapshots. All steps run in a sandboxed environment with a fixed project scaffold and build tools, enabling the agents to execute code, render the result under a mobile viewport, and iterate safely.

\textbf{Stage 1: Deconstruct and recover.}
Given the desktop input (\autoref{fig:pipeline} \circled{a}), the \textit{Semantic Parser} first renders the visualization and aligns the DOM/SVG structure with the rendered layout to identify the chart topology and its components (e.g., axes, legends, marks, annotations) (\autoref{fig:pipeline} \circled{c} and \circled{d}). Next, the \textit{Data Extractor} reconstructs the underlying dataset and key visual specifications (e.g., scale domains and mappings) from the source structure (\autoref{fig:pipeline} \circled{b}) and labels (\autoref{fig:pipeline} \circled{c}), and outputs a structured JSON representation of data (\autoref{fig:pipeline} \circled{e}).

\textbf{Stage 2: Plan the mobile transformation.}
The \textit{Design Planner} takes the recovered data/specs (\autoref{fig:pipeline} \circled{f}) and produces an explicit transformation plan (\autoref{fig:pipeline} \circled{h}).
Guided by the multi-level design space (Section~\ref{sec:design_space}), the plan specifies (i) view-level reorganization (e.g., serialization of multiple views, axis transposition), (ii) reference-frame and component changes (e.g., legend restructuring, externalizing annotations/controls), and (iii) element-level adjustments (e.g., label density, mark sizing), together with concrete implementation directives.
% \todo{Add visual spec here}

\textbf{Stage 3: Implement and verify.}
The frontend engineer implements the plan within the provided mobile scaffold written in \texttt{TypeScript}, producing an executable component (\autoref{fig:pipeline} \circled{i}).
The system then renders this component and captures screenshots under a phone-sized viewport.
The visual critic agent evaluates the output (\autoref{fig:pipeline} \circled{j}) against mobile-first criteria, including data fidelity, plan adherence, text readability, and overall visual quality, and returns targeted feedback (\autoref{fig:pipeline} \circled{k} and \circled{l}). Based on this critique, the workflow loops: code-level issues are routed back to the Engineer, while strategy-level mismatches trigger replanning. The iteration terminates only when the Critic certifies the result as mobile-ready.

Here is an example transform plan example:

\begin{lstlisting}[style=beautiful_listing]
# Vis2Mobile Transformation Plan
## 1. Analysis of Original Visualization & Mobile Challenges
### Visual Spec of Original Visualization (Desktop)
- **Type**: Scatter Plot with Overlayed Confidence Ellipses (Density Estimation).
- **Data**: "Penguins" dataset.
    - **X-Axis**: Flipper Length (mm).
    - **Y-Axis**: Body Mass (g).
    - **Color/Group**: Species (Adelie, Chinstrap, Gentoo).
- **Narrative**: Comparing the correlation between flipper length and body mass across three penguin species. The ellipses highlight the distribution range and correlation strength.
- **Visual Encoding**:
    - **Points**: Individual data entries (Scatter).
    - **Polygons (Ellipses)**: Semi-transparent filled shapes representing statistical deviation.
    - **Legend**: Located on the right side.
### Mobile Challenges (Desktop-on-Mobile Analysis)
1.  **Aspect Ratio Distortion**: The wide aspect ratio (approx 2:1) of the desktop version, when compressed to mobile width, makes the chart extremely short. This squashes the Y-axis range, making vertical separation between points difficult to distinguish.
2.  **Legend Space**: The right-aligned legend consumes approximately 20-25% of the horizontal screen real estate. On mobile, this leaves very little room for the actual data.
3.  **Touch Targets**: The scatter points are relatively small. Hover interactions (standard on desktop) do not exist on mobile.
4.  **Label Readability**: Axis labels and titles may become too small to read if simply scaled down.
5.  **Data Density**: The ellipses overlap significantly. On a small screen, the transparency blending might become muddy without proper color handling.

## 2. Vis2Mobile Design Action Plan

Based on the **Vis2Mobile Design Action Space**, I plan to apply the following actions to transform this visualization.
### L0: Visualization Container
*   **Action: Rescale (Aspect Ratio Adjustment)**
    *   *Reason*: The original "wide" format fails on vertical mobile screens.
    *   *Plan*: Change the container aspect ratio to a taller format (e.g., 1:1 square or 4:5 portrait). This utilizes the vertical scroll behavior of mobile usage and decompresses the Y-axis data.
### L3: Legend Block
*   **Action: Reposition & Transpose**
    *   *Reasoning*: The side legend creates a "Distorted layout" on mobile.
    *   *Plan*: Move the legend to the **top** of the chart (below the main title). Transpose it from a vertical list to a horizontal row of "Chips" or "Badges".
*   **Action: Interaction (Filter)**
    *   *Plan*: Make the legend chips interactive. Tapping a species (e.g., "Adelie") acts as a filter or focus mechanism, dimming the other species to reduce "Overplotting" on the small screen.
### L3/L4: Coordinate System (Axes)
*   **Action: Decimate Ticks (Adjust Ticks)**
    *   *Reason*: High density of tick labels on the X-axis will lead to overlapping text on narrow screens.
    *   *Plan*: Reduce `tickCount` for the X-axis (e.g., max 5 ticks).
*   **Action: Simplify Label**
    *   *Reason*: "Flipper Length (mm)" is long.
    *   *Plan*: Keep the unit but ensure font size is readable (min 12px). If space is tight, move units to the subtitle or a corner label.
### L2: Data Marks (Scatter & Ellipses)
*   **Action: Rescale (Mark Size)**
    *   *Reason*: Small dots are hard to see and tap.
    *   *Plan*: Increase the base size of the scatter circles.
*   **Action: Recompose (Handling Ellipses in Recharts)**
    *   *Constraint*: Recharts standard `Scatter` doesn't support arbitrary filled polygons easily.
    *   *Plan*: Use the `Customized` component or SVG `path` overlays within the Recharts container to render the ellipse data using the extracted coordinates. This ensures the "Confidence/Deviation" narrative is preserved.
### L5: Interaction & Feedback
*   **Action: Reposition (Fix Tooltip Position)**
    *   *Reason*: Standard tooltips follow the cursor. On mobile, the finger obscures the tooltip.
    *   *Plan*: Implement a **"Selected State"**. When a user taps a dot, display the specific data details (Sex, Island, exact mass) in a fixed card at the bottom of the visualization or in a dedicated "Active Data" header space.
*   **Action: Disable Hover / Enable Tap**
    *   *Reason*: Touch interface.
### Premium Aesthetics (UI Layer)
*   **Theme**: Glassmorphism.
    *   Background: Semi-transparent blurred cards for the legend and tooltip area.
    *   Colors: Vivid, accessible palette for the three species (Blue, Orange, Red) ensuring high contrast against the background.
    *   Animation: Smooth transitions when filtering species.
## 3. Data Extraction Strategy
I will extract the data directly from the HTML source `spec` variable.
1.  **Ellipse Data**: Located in `spec.datasets["data-cb82d8ff5c0..."]`.
    *   *Fields*: `order`, `Flipper Length (mm)`, `Body Mass (g)`, `Species`.
    *   *Usage*: This will be converted into three separate arrays (one per species) to draw the SVG paths.
2.  **Scatter Data**: Located in `spec.datasets["data-300f0a3d0..."]`.
    *   *Fields*: `Species`, `Island`, `Beak Length (mm)`, `Beak Depth (mm)`, `Flipper Length (mm)`, `Body Mass (g)`, `Sex`.
    *   *Usage*: Main data source for the scatter plot.
**Note on Data Processing**: The raw ellipse data is a set of points. To render this in Recharts/SVG, I will need to sort them by `order` and construct an SVG path string (`M x1 y1 L x2 y2 ... Z`) for each species.
## 4. Implementation Plan (Step-by-Step)
### Step 1: Data Preparation (`src/data/penguinData.ts`)
*   Create a TS file to export `scatterData` and `ellipseData`.
*   Clean keys (remove spaces/units from keys for easier coding, e.g., `Body Mass (g)` -> `bodyMass`).
*   Group ellipse data by species and sort by `order` to prepare for path generation.
### Step 2: Component Architecture (`src/components/Visualization.tsx`)
*   **Layout Wrapper**: A standard container with padding, utilizing Tailwind for responsive width.
*   **Header**: Title ("Penguin Morphology") and Subtitle ("Body Mass vs Flipper Length").
*   **Controls**: A flex-row of toggle buttons for Species (Legend).
*   **Chart Container**: `ResponsiveContainer` wrapping a `ComposedChart`.
### Step 3: Visualization Implementation (Recharts)
*   **X Axis**: `dataKey="flipperLength"`, type number, domain `['auto', 'auto']` (or specific range based on data to avoid whitespace).
*   **Y Axis**: `dataKey="bodyMass"`, type number.
*   **Ellipses Layer**:
    *   Use Recharts `<Customized />`.
    *   Inside the custom component, iterate through the 3 species.
    *   Use the X and Y scales provided by Recharts to transform data coordinates to pixel coordinates.
    *   Render `<path>` elements with `fillOpacity={0.2}`.
*   **Scatter Layer**:
    *   Use `<Scatter />` component.
    *   Map data to the axes.
    *   Color points based on species.
### Step 4: Mobile Interaction Logic
*   State: `activeSpecies` (for filtering), `selectedPoint` (for tooltip).
*   Interaction:
    *   Tap Scatter Point -> Set `selectedPoint`.
    *   Tap Background -> Clear `selectedPoint`.
*   Feedback:
    *   Render a **"Detail Card"** (Glassmorphism style) floating at the bottom or fixed below the header when a point is selected.
### Step 5: Styling & Polish
*   Apply Tailwind classes for typography (Inter font, legible sizes).
*   Add subtle animations using `framer-motion` (optional) or CSS transitions for the filter chips.
*   Ensure axis lines are minimal (`stroke-gray-200`) and grid lines are dashed and subtle.
This plan moves the visualization from a static, desktop-centric analysis tool to an interactive, mobile-friendly exploration interface.
    
\end{lstlisting}

% \subsection{Workflow Execution}
% The system orchestration follows a dynamic pipeline:
% \begin{enumerate}
%     \item \textbf{Initialization:} The system sets up a sandboxed environment with necessary tools (e.g., \texttt{bun}, \texttt{uv}) for safe code execution.
%     \item \textbf{Planning Phase:} The Parser and Planner agents collaborate to produce the \texttt{transform-plan.md}.
%     \item \textbf{Coding Phase:} The Engineer agent implements the plan, creating the \texttt{src/components/Visualization.tsx} file and associated stylesheets.
%     \item \textbf{Verification Phase:} The system executes a rendering script (e.g., \texttt{screenshot.py}) to visualize the output. The Critic agent reviews the result against the "mobile-first" criteria.
%     \item \textbf{Refinement Loop:} Upon receiving critique, the Engineer agent modifies the code. This cycle continues until the Critic certifies the visualization as "Mobile-Ready."
% \end{enumerate}

% This structured approach ensures that Proteus does not simply "fit" content to a screen, but actively "re-authors" it for optimal human consumption.

\begin{figure*}
    \centering
    \includegraphics[width=\textwidth]{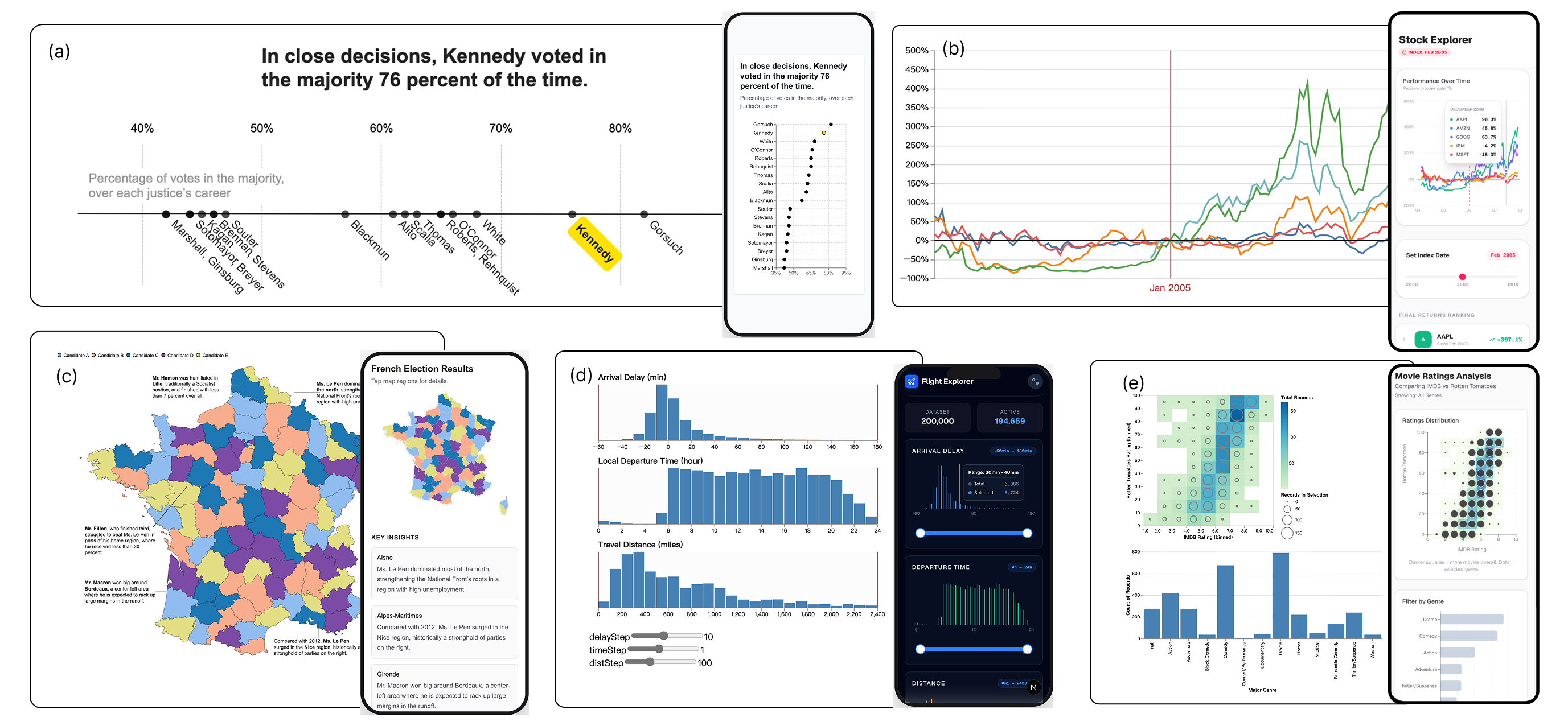}
    \caption{Case studies of \toolname{} on five real-world web visualizations. For each case, the left shows the original desktop design and the right shows the corresponding mobile visualization generated by \toolname{}.}
    \Description{A series of examples showing the transformation of real-world desktop visualizations into mobile formats, with each case contrasting the original design and the adapted mobile result produced by the system.}
    \label{fig:cases}
\end{figure*}

\section{Case Studies}

As shown in \autoref{fig:cases}, we collected a set of online visualization examples from the web and used them as cases to run our system. These examples cover common chart types, including line charts, bar charts, maps, and scatter plots. In each case, the left side shows the original desktop visualization, and the right side shows the mobile visualization automatically generated by our method.

\subsection{Single-View Visualizations}

The original visualization\footnote{https://see-mike-out.github.io/cicero-supplemental/} in \autoref{fig:cases} (a) shows, for each U.S. Supreme Court justice, the percentage of votes in closely decided cases in which the justice sided with the majority. The horizontal axis encodes “percentage of votes in the majority,” and each dot corresponds to a justice. The chart particularly highlights Kennedy, whose percentage is 76\%, to emphasize that he tends to align with the majority in such close decisions.
The original chart is a very wide 1D scatterplot designed for desktop displays. Our method keeps the scatterplot as the main visual encoding on mobile, but redesigns the layout to better fit a small, vertically oriented screen.
If we directly scale the original chart down to a phone, the text becomes too small and there is a large amount of unused vertical space, which harms readability.
To address this, \toolname{} preserves the original horizontal data mapping but reorganize the items along the vertical dimension on mobile. Instead of spreading the justices horizontally, we stack them vertically, which makes more efficient use of the vertical screen space and reduces wasted whitespace.
This new layout preserves the information conveyed by the original visualization while making the structure more compact and readable on mobile. In addition, we reorganize the title and descriptive text: the original in-chart annotation (\textit{``percentage of votes in the majority, over each justice's career''}) is moved into a subtitle, which reduces text clutter inside the plot area and provides necessary context before the user reads the chart.

Similarly, \autoref{fig:cases} (c) is a nearly square choropleth map of France\footnote{https://see-mike-out.github.io/cicero-supplemental/}, showing which presidential candidate received the most votes in each d\'epartement. Different colors encode different candidates, and the original desktop design uses several long textual annotations placed around the map to highlight regional patterns (e.g., where specific candidates performed especially well or poorly).
On mobile, our method preserves the map itself but adapts the surrounding layout to the vertical screen. If we simply scaled the original chart down, the square map would leave a large unused area below it, and the dense in-map annotations would quickly become unreadable. Instead, we externalize these textual annotations into a “Key insights” panel placed below the map. Each insight is presented as a separate, scrollable card that summarizes an important regional pattern, such as strongholds of particular candidates or notable shifts compared with previous elections.
In addition, the mobile map supports common interactions such as zooming, panning, and tooltips on tap. These interactions allow users to explore specific regions in detail despite the limited screen size, complementing the summarized insights and helping readers investigate areas of interest on demand.

External components can not only present textual information (as in cases (a) and (c)), but also act as interactive controls that adapt and extend the original interaction for mobile devices.
In case (b), the original desktop visualization\footnote{https://vega.github.io/vega/examples/stock-index-chart/} already supports setting the stock index date.
It is implemented through mouse hover: when the user hovers over a vertical reference line at a given time point, the chart recomputes returns relative to that index date.
However, this hover-based interaction does not translate well to mobile. On the one hand, the limited ``hover'' capability on touch devices is better reserved for tooltips, which are crucial for inspecting exact values. On the other hand, dragging directly on the chart with a finger can occlude the lines and hinder readability. To resolve this conflict, we externalize the “set index date” interaction into a separate control placed below the chart.
In our mobile design, users adjust the index date via a dedicated “Set Index Date” slider, while the line chart above updates to show returns relative to the selected date. At the same time, touch interactions on the chart are dedicated solely to triggering tooltips for precise value inspection. In this way, the original index-setting functionality is preserved, but reimplemented in a manner that better matches mobile interaction habits and avoids overloading the hover semantics.

\subsection{Multi-View Visualizations}

When a visualization comprises multiple coordinated charts, a mobile redesign must not only resize individual views but also reorganize the overall layout to preserve readability and coordination. 
Moreover, these views often support cross-filtering or cross-highlighting, where selections in one chart affect the others. Such multi-view coordination is typically already encoded in the original desktop visualization; otherwise, the webpage would not be able to support cross-linking in the first place. Our approach extracts these inter-view relationships and reassembles them into a mobile-friendly structure that maintains the original analytical workflow.

As shown in \autoref{fig:cases} (d), our method preserves the original filtering widgets\footnote{https://vega.github.io/vega/examples/crossfilter-flights/} and keeps slider-based interaction as the primary mechanism for refining the dataset. However, instead of placing all controls in a dense block (as on desktop), we reposition each slider to sit directly with its corresponding view, creating a one-to-one mapping between control and chart. In addition, we extend the slider design: while the desktop version only supports setting an upper bound, our mobile version provides a range slider that allows users to specify both lower and upper bounds. The filtered results are immediately reflected across all coordinated views, and the interface explicitly reports the number of selected items, making the cross-view effect of filtering visible and easy to verify.

In \autoref{fig:cases} (e), the original desktop design\footnote{\url{https://altair-viz.github.io/gallery/interactive_cross_highlight.html}} consists of two coordinated views: a binned ratings distribution (2D grid with circles) on top and a genre bar chart below. The desktop visualization supports cross-view linking: selecting a genre in the bar chart highlights the corresponding subset of movies in the ratings distribution, enabling users to compare the selected genre against the overall population shown in the background.
Our mobile redesign preserves this coordination but reimplements the interaction in a touch-friendly form. Users tap a bar in the lower chart to select a genre, and the upper distribution view immediately updates to highlight the selected movies while keeping the remaining records as contextual background. To better fit the narrow, vertical screen and improve tap precision, we also reorient the genre bar chart from vertical columns to horizontal bars, which provides more space for category labels and creates larger, easier-to-target touch areas.
% \section{User Study}

\section{Evaluation}
\label{sec:evaluation}

To validate the effectiveness of \toolname{}, we conducted a comparative user study focusing on the quality of the generated mobile visualizations. Our evaluation aims to answer the following research question: \textit{Does the integration of a Multi-level Design Space significantly improve the readability, aesthetics, and data fidelity of mobile adaptations compared to generic LLM-based approaches?}
\revision{We note that we do not include direct system-level comparisons with MobileVisFixer~\cite{wu_mobilevisfixer_2020} or Cicero~\cite{cicero2022}. This is because these prior works are not directly comparable to our task setting. MobileVisFixer focuses on a more restricted class of visualizations, primarily single Cartesian charts with linear or discrete scales, whereas \toolname{} targets a broader range of visualization specifications.
Cicero is a visualization grammar rather than an end-to-end system for automatic mobile transformation.
Therefore, instead of reporting potentially misleading direct comparisons, we focus our evaluation on comparisons with generic LLM-based baselines.}

\subsection{Benchmark Dataset Construction}

To rigorously evaluate our framework, we constructed a diverse benchmark dataset comprising 67 representative desktop visualizations.
We exclude examples that are near-duplicates in structure and style.
These examples were curated from official galleries (e.g., the Vega examples gallery~\cite{vega-gallery}, the Vega-Lite examples gallery~\cite{vegalite-gallery}, the Altair example gallery~\cite{altair-gallery}, and the D3 examples~\cite{d3-gallery}) to ensure variety.
This collection spans a broad spectrum of data density and structural complexity, from simple statistical charts (e.g., bar, line, area, and scatter plots) to views with many visual elements or composite layouts.
As a result, the dataset poses diverse challenges for mobile adaptation, including dense encodings, implementation methods, and multi-view composition.

\subsection{Comparative Baselines}
A straightforward baseline is to directly prompt a single LLM to generate mobile visualization code in one shot. However, in our preliminary tests, this approach exhibited a very low correctness rate: in most cases, the produced code failed to render.
Since such failures prevent meaningful comparison on visual quality, we consider this setting too weak and ultimately unfair as a baseline.
To ensure a fair comparison with our agentic system, we therefore implemented a stronger baseline using the same Multi-agent architecture (Planner + Engineer + Critic) but removed the knowledge of the Multi-level Design Space. This baseline has access to the same toolchain, coding environment, and execution feedback as \toolname{}, and can iteratively revise code through agent collaboration.
We use the latest \textbf{Gemini 3.0 Pro Preview}~\cite{gemini3pro} API as a high-capacity LLM backbone for all agents in both \toolname{} and this baseline. 
For a single visualization, the end-to-end adaptation process typically completes within 10 minutes (about 2–10 minutes in our experiments).

\subsection{Metrics and Procedure}
We recruited {12} participants, all with prior experience in data visualization work, to evaluate the outputs. The study followed a within-subjects design. For each case in the dataset, participants were shown the original desktop chart and two mobile adaptations: one generated by \toolname{} and one by the baseline.
Each participant completed all cases in approximately 60–90 minutes.
Participants evaluated the adaptations based on the following five dimensions. Note that D0 serves as a binary prerequisite, while D1--D4 are rated on a 7-point Likert scale:

\begin{itemize}
    \item \textbf{D0. Execution Completeness:}
    Before rating quality, participants first judged whether the system successfully generated a renderable visualization. A trial was marked as failure if the output contained syntax errors, produced a blank screen, or only showed an obviously incomplete rendering.  Otherwise, the visualization was directly assigned a score of 0 on D1--D4 without further evaluation.

    \item \textbf{D1. Data Fidelity:} \textit{Does the mobile chart accurately reflect the data of the original desktop chart without hallucination or critical loss?} (1 = Severe Distortion, 7 = Perfect Fidelity).

    \item \textbf{D2. Perceptual Readability:} \textit{Is the text legible? Are visual marks clearly distinguishable without overlapping?} (1 = Unreadable, 7 = Highly Readable).

    \item \textbf{D3. Interaction Reasonableness:} \textit{Is the information easy to consume on a phone? Is the interaction design reasonable and practically useful?} (1 = Useless, 7 = Very Useful).

    \item \textbf{D3. Visual Aesthetics:} \textit{Is the layout balanced and visually pleasing?} (1 = Cluttered/Broken, 7 = Professional).
    % For instance, in \autoref{fig:comparison}(b2) (our result) and \autoref{fig:comparison}(b3) (baseline), if participants perceive (b2) as having a more polished visual style, they would assign a higher aesthetics score to (b2).

\end{itemize}
We acknowledge that these metrics are not purely objective and inevitably involve subjective judgment. To mitigate potential bias, we adopted the following randomization and blinding strategies:

\begin{itemize}
    \item \textbf{Randomized case order.} For each participant, the sequence of visualizations was randomly shuffled, reducing order effects and learning effects across tasks.
    \item \textbf{Randomized side assignment.} For each case, the positions of the two mobile adaptations (left vs.\ right) were randomized, and participants were not informed which side corresponded to \toolname{} or the baseline. This prevents systematic preference due to side bias and ensures that ratings reflect perceived quality rather than system identity.
\end{itemize}

% Together, these procedures ensure that the collected scores provide a fair and unbiased comparison between our method and the baseline under identical viewing and interaction conditions.

\begin{figure*}[htb]
    \centering
    \includegraphics[width=\textwidth]{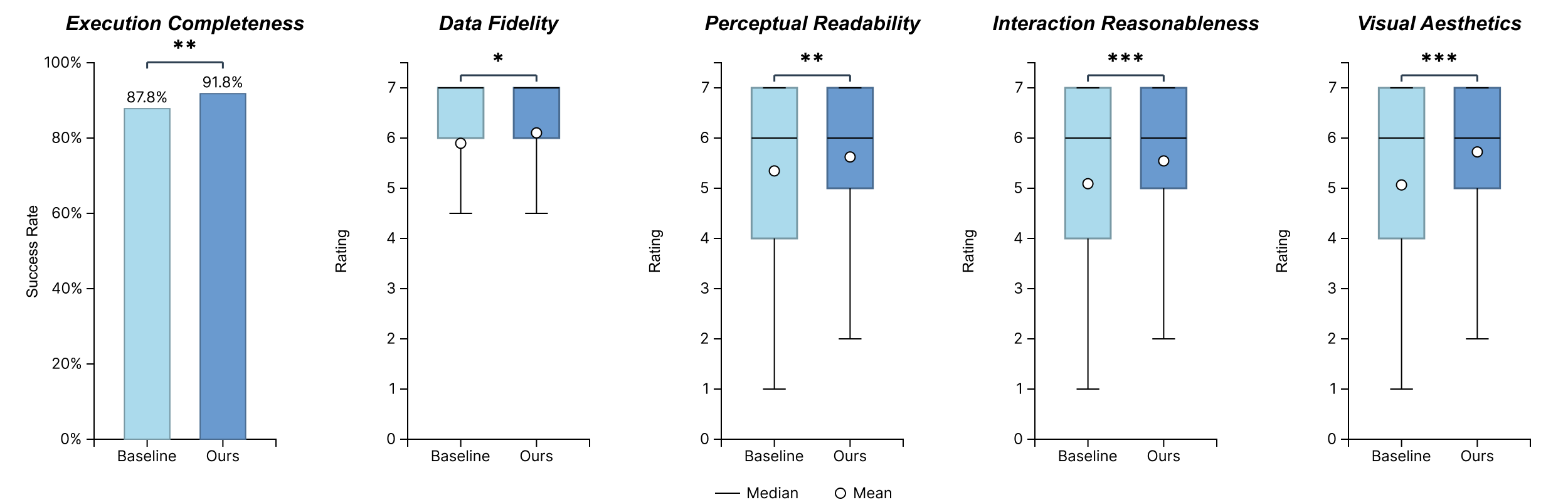}
    \caption{Results of the user study comparing \toolname{} with the baseline across five evaluation dimensions. 
    The leftmost chart shows the success rate of rendering (D0), while the remaining boxplots report participant ratings for data fidelity (D1), perceptual readability (D2), interaction reasonableness (D3) and visual aesthetics (D4).
    We use the Wilcoxon signed-rank test to assess statistical significance; the number of asterisks above each pair denotes the significance level
    (\textbf{*} $p < 0.05$, \textbf{**} $p < 0.01$, \textbf{***} $p < 0.001$). 
    Overall, \toolname{} significantly outperforms the baseline on all dimensions.
    }
    \Description{Charts summarizing results from a user study comparing Proteus with a baseline approach, including rendering success rates and user ratings across multiple criteria, with indicators showing statistically significant differences favoring Proteus.}
    \label{fig:user_study_results}
\end{figure*}

\subsection{Results}
We aggregated the user ratings and performed statistical analysis using the Wilcoxon signed-rank test to determine significance.

Across all benchmark visualizations, \toolname{} successfully completed the end-to-end adaptation pipeline on 91.8\% of cases, as judged by participants' D0 ratings, while the baseline achieved a lower render success rate of 87.8\%. 
For each failed case, D0 was marked as failure and the corresponding D1--D4 scores were set to 0.
We revisited the originally failed D0 case and found that the failure did not recur in a repeated run, suggesting that it may have resulted from transient system instability rather than a persistent rendering issue.
In total, 12 participants each evaluated 67 visualization pairs, providing ratings on five dimensions (D0–D4), which resulted in 804 paired comparisons per dimension. 
We applied the Wilcoxon signed-rank test because the rating distributions deviated from normality.

As summarized in Fig.~\ref{fig:user_study_results}, \toolname{} significantly outperforms the baseline across all evaluation dimensions. In terms of data fidelity (D1) and text readability (D2), our method achieves higher ratings with statistical significance ($p < 0.05$). For interaction reasonableness (D3) and visual aesthetics (D4), the improvements are even more pronounced, with \toolname{} showing highly significant advantages ($p < 0.001$).

\begin{figure*}[ht]
    \centering
    \includegraphics[width=\textwidth]{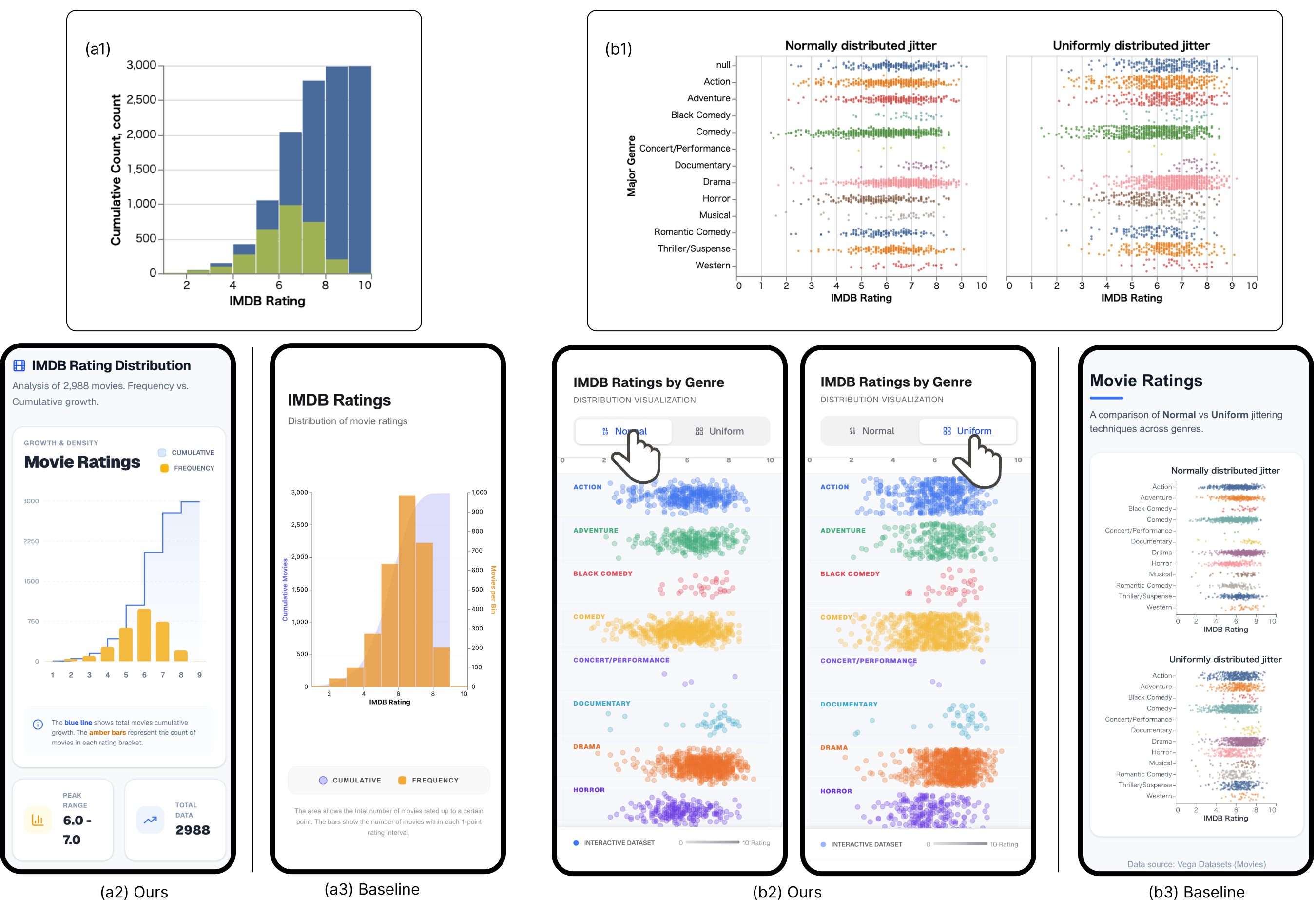}
    \caption{Comparison between \toolname{} and the baseline on two representative cases. 
    (a1) and (b1) show the original desktop visualizations. 
    (a2) and (b2) present the mobile adaptations generated by \toolname{}, while (a3) and (b3) show the corresponding mobile versions produced by the baseline. 
    In (a), our method better preserves the underlying data distribution, whereas the baseline produces incorrect values even though the overall trend appears similar. 
    In (b), \toolname{} provides a more readable layout and a more reasonable interaction (e.g., switching between \textit{Normal} and \textit{Uniform} jitter).}
    \Description{Comparative examples of two visualization cases highlighting differences between Proteus and a baseline method, emphasizing improvements in data accuracy, layout clarity, and interaction design in the system’s outputs.}
    \label{fig:comparison}
\end{figure*}

These quantitative results are consistent with participants’ qualitative judgments on specific cases. For example, in \autoref{fig:comparison} (a2) (our result) and \autoref{fig:comparison} (a3) (baseline), the overall trend in (a3) may look roughly similar at first glance, but the actual encoded values deviate from the original data; consequently, (a3) receives a much lower data fidelity score. Similarly, in Fig.~\ref{fig:comparison} (b2), our mobile version allows users to switch between \textit{normal} and \textit{uniform} modes via an external icon with a smooth animated transition. Compared with a version that provides little or no meaningful interaction, this more reasonable interaction design is judged as more useful for real mobile usage, leading to higher D3 scores.

\section{Discussion}

% 一、扩展到 Bitmap 图像。
% 我们在当前的任务中，选择了在线可视化的通常的形式，即通过矢量形式（HTML、SVG）等方式的可视化来出发。
% 这类可视化的数据是从矢量的结构中可以恢复的。
% 而这也是最常用的可视化。

% 覆盖性，我们展示了我们的方法在常用的可视化中，基本上可以很好的恢复出对应的可视化，而在较复杂的可视化中。
% 由于原本的可视化的规模较大，需要对数据的处理，以及写代码的复杂度也会上升。

% 二、支持的范围和时间。
%  对于简单的可视化，整个流程可以很快的。
% 我们从 d3 的gallery 和 vega-lite。发现这些可视化大都可以被我们的方法所支持。
% 只是对于复杂的可视化。

% 三、
In this work, we presented a generative framework for automating the adaptation of visualizations from desktop to mobile environments. Our results demonstrate the efficacy of mapping design requirements to a multi-level design space, but several limitations and directions for future research remain.

\subsection{Complexity and the Long Tail of Design}

% Our evaluation on standard galleries (e.g., D3 and Vega-Lite examples) suggests that the framework handles canonical chart types with high fidelity and reasonable efficiency. Nonetheless, the broader visualization ecosystem includes a long tail of bespoke, highly customized, and artistic designs that diverge from conventional grammars.

% For such complex cases, two main challenges emerge. First, inference latency tends to increase with the structural complexity of the SVG or DOM representation, as larger and more intricate scene graphs require more tokens for the LLM to process. Second, many bespoke visualizations rely on custom layout logic (e.g., force-directed simulations, hierarchical packing, or domain-specific encodings) that do not decompose cleanly into our predefined atomic operators. While the system is often able to produce a functional mobile variant, preserving subtle aesthetic and stylistic qualities remains difficult. Addressing this long tail may require domain-specific model tuning and richer, higher-level operator libraries tailored to particular visualization genres.

Our evaluation on standard galleries, such as D3 and Vega-Lite examples, suggests that the framework handles canonical chart types with high fidelity and reasonable efficiency. Nonetheless, the broader visualization ecosystem includes a long tail of bespoke, highly customized, and artistic designs that diverge from conventional grammars.

For such complex cases, two main challenges arise. First, inference latency tends to increase with the structural complexity of the SVG or DOM representation, since larger and more intricate scene graphs require the LLM to process more tokens. Second, many bespoke visualizations rely on custom layout logic, such as force-directed simulations, hierarchical packing, or domain-specific encodings, which do not decompose cleanly into our predefined atomic operators. Although the system can often produce a functional mobile variant, preserving subtle aesthetic and stylistic qualities remains difficult. Addressing this long tail may require domain-specific model adaptation and richer high-level operator libraries tailored to particular visualization genres.

\revision{In addition, our current evaluation primarily examines visual fidelity and perceived comparability. It does not directly assess whether the adapted visualizations support the same analytic tasks or lead users to the same conclusions as the original designs. This limitation may be especially important for bespoke visualizations, where the effect of adaptation can vary across task types. A task-oriented evaluation of analytic equivalence is an important direction for future work.}

\subsection{The Paradigm of Automated Consumption}

Unlike mixed-initiative tools for visualization authoring (for example, recommendation systems that assist designers during creation), our framework targets the consumption side of the visualization pipeline. In the scenarios we consider, such as reading data-rich articles on mobile devices, users primarily assume the role of readers rather than editors. They typically lack the time, expertise, or interaction affordances to engage in iterative design refinement or to explicitly approve adaptation choices.

This setting imposes a stronger requirement for zero-intervention robustness: the system must operate as an autonomous design proxy, resolving design ambiguities (for example, whether to transpose an axis or introduce scrolling) without human confirmation.
\revision{A critical component in achieving this autonomy is the critic agent, which evaluates intermediate designs and guides the iterative refinement process.
In our workflow, most cases involve multiple rounds of refinement, and we observe that without the critic, the system frequently fails to converge on functional mobile variants.
This underscores that the critic is not a necessary part of the adaptation pipeline.}
Although this zero-intervention constraint limits the degree of fine-grained, human-curated optimization that is possible, it enables a pervasive reading-oriented usage mode for visualization and lowers the barrier to accessing complex data on handheld devices, where manual adjustment would be impractical.

\begin{figure}[ht]
    \centering
    \includegraphics[width=\linewidth]{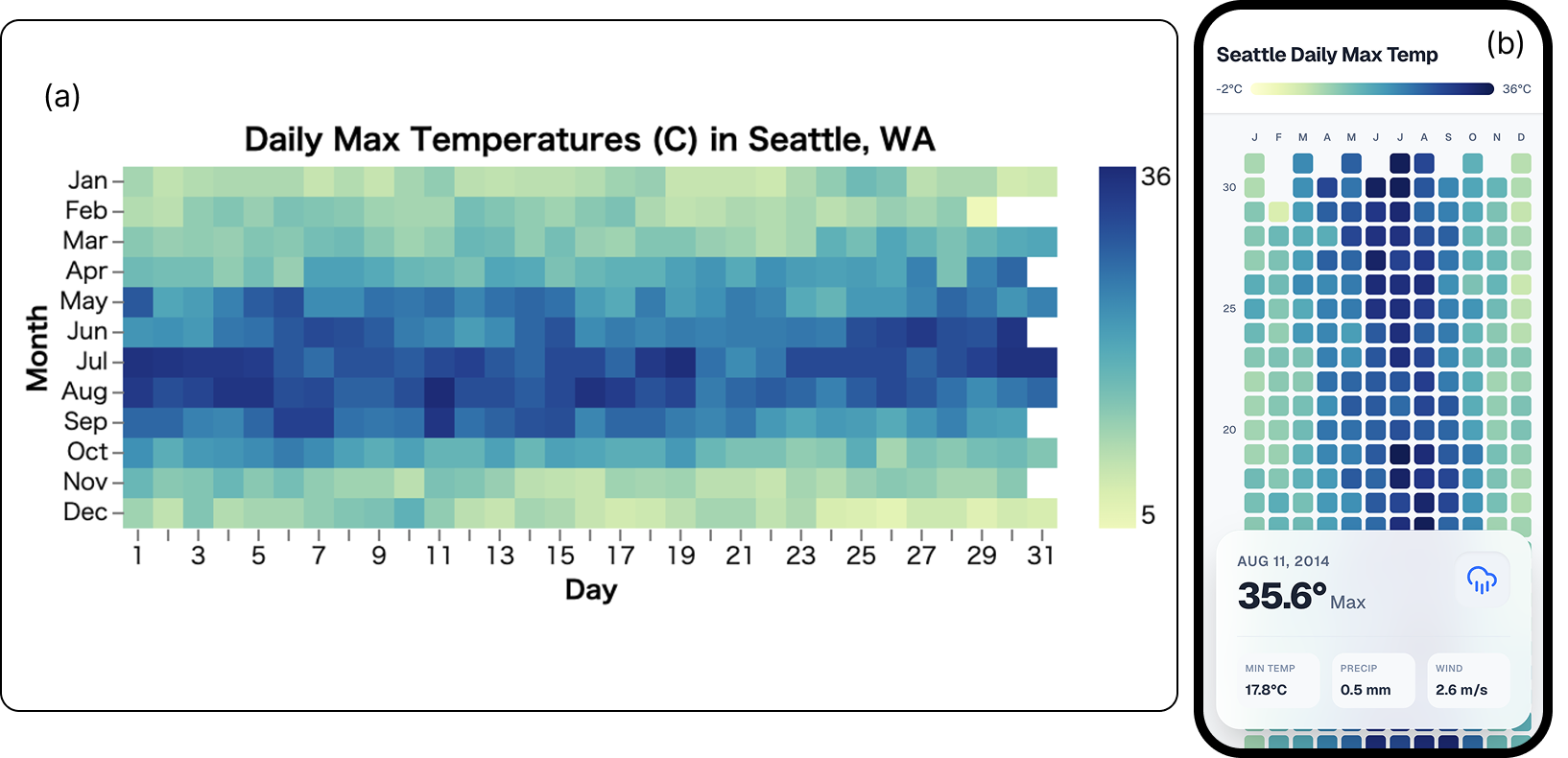}
    \caption{Style preservation during mobile adaptation. (a) The original desktop visualization. (b) The adapted mobile visualization generated under a constraint that preserves the original color scheme more closely.
    }
    \Description{An illustration of style preservation during adaptation, comparing an original visualization with a mobile version that maintains a similar color scheme under constrained conditions.}
    \label{fig:keep_style}
\end{figure}

\revision{Future work could explore lightweight preference mechanisms that lie between full automation and full authoring control, for example by allowing users to specify high-level defaults, such as preferring scrolling over pagination or favoring reduced clutter over maximal detail, without requiring them to manipulate visualization parameters directly.
One direction is to expose style-preservation preferences at a high level.
For example, although our default adaptation strategy may revise color choices to improve readability and visual clarity on small screens, users can optionally request stronger preservation of the original color scheme. In a rerun of one representative example (as shown in \autoref{fig:keep_style}) under such a constraint, we found that the adapted mobile visualization could still achieve good usability while more closely matching the source style.}
% Future work could explore lightweight preference mechanisms that lie between full automation and full authoring control, for example allowing users to specify high-level defaults (such as preferring scrolling over pagination or favoring reduced clutter over maximal detail) without requiring them to manipulate visualization parameters directly.

\subsection{Generalizability to Raster Graphics}

Our current implementation operates primarily on vector-based specifications (e.g., SVG), where the underlying data and structural semantics are directly accessible. This focus aligns with the dominant paradigm of modern web-based visualization (e.g., D3, Vega-Lite). However, a substantial portion of existing visualizations, especially legacy content, remains available only as static raster images, where data and marks are encoded purely at the pixel level.

Extending our framework to raster graphics would require an additional upstream component for reverse engineering. Such a component would employ computer vision and OCR techniques to reconstruct a scene graph and recover an approximate underlying data table. Although recent multimodal language models show encouraging progress in chart understanding, accurate, fine-grained deconstruction of arbitrary charts from pixels is still an open challenge. Future work could integrate our adaptation engine with chart-to-code models, forming a pipeline that reconstructs, adapts, and re-renders static bitmap visualizations as responsive mobile components.

\section{Conclusion and Future Work}
\label{sec:conclusion}

\revision{We addressed the gap between desktop-oriented visualization authoring and mobile consumption, where simple resizing often harms readability and interaction. We argued that effective mobile adaptation requires more than geometric resizing and should instead revise the visualization content in a meaning-preserving way.
To this end, we proposed a multi-level design space that organizes adaptation as a hierarchy of decisions, from global structure, through reference elements, to individual visual elements. Guided by this framework, we introduced \toolname{}, an LLM-driven multi-agent system that parses heterogeneous web charts, plans semantic transformations, and generates executable mobile-optimized views.
A controlled user study on real-world visualizations shows that \toolname{} outperforms a strong multi-agent LLM baseline in rendering success and four qualitative dimensions, including data fidelity, readability, interaction quality, and visual aesthetics.
In the future, we plan to extend \toolname{} by supporting rapid end-user customization of mobile adaptations and broadening input coverage from vector-based web charts to raster-only visualizations.
}

% We addressed the authoring–consumption gap that arises when desktop-oriented visualizations are viewed on mobile devices, where simple scaling often breaks readability and interaction. We argued that mobile-first experiences require automated semantic re-authoring rather than flat geometric heuristics.
% To this end, we proposed a multi-level design space that models adaptation as hierarchical constraint propagation from global topology, through reference frames, down to individual visual elements. Guided by this framework, we introduced \toolname{}, an LLM-driven multi-agent system that parses heterogeneous web charts, plans semantic transformations, and generates executable mobile-optimized views.
% A controlled user study on real-world visualizations shows that \toolname{} outperforms a strong multi-agent LLM baseline in rendering success and 4 qualitative dimensions, including data fidelity, readability, interaction quality, and visual aesthetic. 
% In the future, we plan to extend \toolname{} by supporting rapid end-user customization of mobile adaptations and broadening input coverage from vector-based web charts to raster-only visualizations.
% Future work includes extending our framework to cross-modal adaptation (e.g., audio or haptics) and personalizing adaptations to user expertise and goals.

\end{CJK}

\begin{acks}
This project is supported by the Ministry of Education, Singapore, under its Academic Research Fund Tier 2 (Proposal ID: T2EP202220 049), and by the Start Up Grant awarded to Yong Wang. Any opinions, findings and conclusions, or recommendations expressed in this material are those of the author(s) and do not reflect the views of the Ministry of Education, Singapore.
\end{acks}

\bibliographystyle{ACM-Reference-Format}
\bibliography{main}

\appendix

\end{document}